\documentclass[twocolumn,5p,times,final]{elsarticle}
\usepackage{graphicx,amssymb}
\journal{J. Magn. Magn. Mater.}

\begin{document}

\begin{frontmatter}

\title{Ground-state phase diagram and magnetization process of the exactly solved mixed spin-(1,1/2) Ising diamond chain}
\author[UPJS,ICMP]{Bohdan Lisnyi\fnref{ack1}}
\ead{lisnyj@icmp.lviv.ua}
\fntext[ack1]{B.L. acknowledges the financial support provided by the National Scholarship Programme of the Slovak Republic for the Support of Mobility of
Students, PhD Students, University Teachers, Researchers and Artists.}
\author[UPJS]{Jozef Stre\v{c}ka\fnref{ack2}}
\ead{jozef.strecka@upjs.sk}
\fntext[ack2]{J.S. acknowledges the financial support provided by the grant of The Ministry of Education, Science, Research and Sport of the Slovak Republic 
under the contract No. VEGA 1/0234/12 by the ERDF EU (European Union European regional development fond) grant provided under the contract No. ITMS26220120005 (activity 3.2).}
\address[UPJS]{Department of Theoretical Physics and Astrophysics, Faculty of Science, 
P. J. \v{S}af\'{a}rik University, Park Angelinum 9, 040 01 Ko\v{s}ice, Slovak Republic}
\address[ICMP]{Institute for Condensed Matter Physics, National Academy of Sciences of Ukraine, 1 Svientsitskii Street, 79011 L'viv, Ukraine}

\begin{abstract}
The ground state and magnetization process of the mixed spin-(1,1/2) Ising diamond chain is exactly solved by employing the generalized decoration-iteration mapping transformation and the transfer-matrix method. The decoration-iteration transformation is first used in order to establish a rigorous mapping equivalence with the corresponding spin-1 Blume-Emery-Griffiths chain in a non-zero magnetic field, which is subsequently exactly treated within the framework of the transfer-matrix technique. It is shown that the ground-state phase diagram includes just four different ground states and the low-temperature magnetization curve may exhibit an intermediate plateau precisely at one half of the saturation magnetization. Our rigorous results disprove recent Monte Carlo simulations of Zihua Xin \textit{et al.} [Z. Xin, S. Chen, C. Zhang, J. Magn. Magn. Mater. 324 (2012) 3704], which imply an existence of the other magnetization plateaus at 0.283 and 0.426 of the saturation magnetization.
\end{abstract}


\end{frontmatter}

\section{Introduction}

During the last few years, a considerable research interest has been devoted to the frustrated magnetism of diamond spin chains that was originally initiated by the effort to clarify several unusual magnetic features of the natural mineral azurite Cu$_3$(CO$_3$)$_2$(OH)$_2$ like for instance a presence of the one-third magnetization plateau in the low-temperature magnetization process \cite{kik04,kik05,kik06,jes11,hon11}. To provide a comprehensive description of the overall magnetic behaviour of the azurite, which is basically affected by a mutual interplay between the geometric spin frustration and quantum fluctuations, one has to employ a rather sophisticated combination of the first-principle density-functional calculations supplemented with the extensive numerical calculations \cite{jes11,hon11}. However, it is worthy of notice that some important vestiges of this intriguing magnetic behaviour can be traced back also from the relevant behaviour of the exactly tractable spin-1/2 Ising diamond chain \cite{val08}. 

On the other hand, the exactly solvable mixed-spin Ising diamond chains have received much less attention so far. To the best of our knowledge, the mixed spin-(1/2,1) Ising chains with the spin-1/2 nodal atoms and the spin-1 interstitial (decorating) atoms have been just marginally investigated as special liming cases of the exactly solved mixed-spin Ising-Heisenberg diamond chains \cite{jas04,can06,roj11}. Quite recently, Zihua Xin \textit{et al.} \cite{xin12} have studied another version of the mixed spin-(1,1/2) Ising diamond chain with the spin-1 nodal atoms and the spin-1/2 interstitial atoms within the framework of numerical Monte Carlo simulations. The main purpose of this work is to provide the exact analytical solution for this mixed-spin Ising diamond chain, which will convincingly contradict a presence of the striking magnetization plateaus at 0.283 and 0.426 of the saturation magnetization theoretically predicted in Ref. \cite{xin12}.

This paper is organized as follows. The model and basic steps of the exact method will be clarified in Sec. \ref{model}. The most interesting results for the ground-state phase diagrams and magnetization process of the symmetric as well as asymmetric mixed-spin Ising diamond chain will be widely discussed in Sec. \ref{result}. Finally, our paper will end up with several concluding remarks and future outlooks mentioned in Sec. \ref{conclusion}.

\section{Model and its exact solution}
\label{model}

\begin{figure}
\begin{center}
\includegraphics[width=0.8\columnwidth]{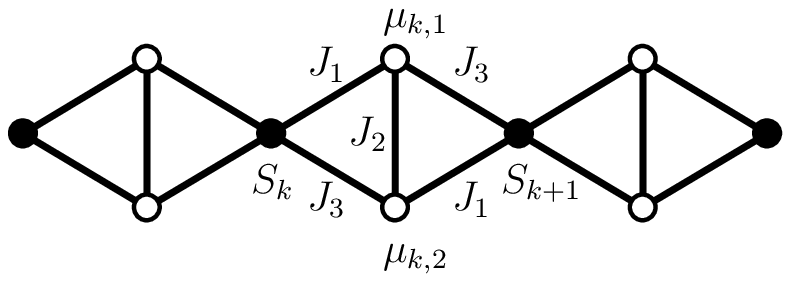}
\end{center}
\vspace{-0.5cm}
\caption{A fragment from the mixed-spin Ising diamond chain. The nodal ($S_{k}$, $S_{k+1}$) and interstitial ($\mu_{k,1}$, $\mu_{k,2}$) Ising spins belonging to the $k$th primitive unit cell are marked. The Ising interactions $J_1$ and $J_3$ along sides of the diamond unit cell are equal (different) in the symmetric (asymmetric) diamond chain.}
\label{fig1}
\end{figure}

Let us begin by considering the mixed spin-(1,1/2) Ising diamond chain in a presence of the external magnetic field. The magnetic structure of the investigated model system is schematically illustrated in Fig.~\ref{fig1} together with its primitive unit cell. As one can see, the primitive unit cell in a shape of diamond spin cluster involves two nodal Ising spins $S_{k}$ and $S_{k+1}$ along with two interstitial Ising spins $\mu_{k,1}$ and $\mu_{k,2}$. The total Hamiltonian for the mixed spin-(1,1/2) Ising diamond chain in a presence of the external magnetic field
reads
\begin{eqnarray}
{\cal H} &=&J_1 \sum_{k=1}^{N} S_{k} (\mu_{k,1} + \mu_{k-1,2}) + J_2 \sum_{k=1}^{N} \mu_{k,1} \mu_{k,2}  \nonumber \\
&& + J_3 \sum_{k=1}^{N} S_{k} (\mu_{k-1,1} + \mu_{k,2}) - h \sum_{k=1}^{N} (S_{k} + \mu_{k,1} + \mu_{k,2}),
\label{htot}
\end{eqnarray}
where the nodal Ising spins $S_k = \pm 1, 0$, the interstitial Ising spins $\mu_{k,\alpha} = \pm 1/2$ ($\alpha = 1,2$) and the periodic boundary conditions $\mu_{0,1} \equiv \mu_{N,1}, \mu_{0,2} \equiv \mu_{N,2}$ are implied for convenience. The interaction constants $J_1$ and $J_3$ label the nearest-neighbour interactions between the nodal and interstitial Ising spins along sides of the primitive diamond unit cell, while the interaction term $J_2$ accounts for the diagonal interaction between the nearest-neighbour interstitial spins from the same primitive cell. Finally, the Zeeman's term $h$ determines the magnetostatic energy of the nodal and interstitial Ising spins in the external magnetic field.

For  further manipulations, it is quite advisable to rewrite the total Hamiltonian (\ref{htot}) as a sum over cell Hamiltonians
\begin{equation}
{\cal H} = \sum_{k=1}^N {\cal H}_k,
\label{hk}
\end{equation}
whereas the cell Hamiltonian ${\cal H}_k$ involves all the interaction terms of the $k$th diamond unit cell
\begin{eqnarray}
{\cal H}_k &=& J_1 (\mu_{k,1} S_{k} + \mu_{k,2} S_{k+1}) + J_2 \mu_{k,1} \mu_{k,2}
+ J_3 (\mu_{k,1} S_{k+1} + \mu_{k,2} S_{k})  \nonumber \\
&& - h (\mu_{k,1} + \mu_{k,2}) - \frac{h}{2} (S_{k} + S_{k+1}).
\label{cell}
\end{eqnarray}
The factor $\frac{1}{2}$ by the last term of Eq.~(\ref{cell}) avoids a double counting of the Zeeman's term for the nodal Ising spins, which is equally split into two different cell Hamiltonians including one and the same nodal Ising spin. The partition function of the mixed spin-(1,1/2) Ising diamond chain can be written in this form
\begin{eqnarray}
{\cal Z} = \sum_{\{S_k \}} \prod_{k=1}^N \sum_{\mu_{k,1}} \sum_{\mu_{k,2}} \exp (-\beta {\cal H}_k)
= \sum_{\{S_k \}} \prod_{k=1}^N {\cal Z}_k,
\label{pfd}
\end{eqnarray}
where $\beta=1/(k_{\rm B} T)$, $k_{\rm B}$ is the Boltzmann's constant, $T$ is the absolute temperature and the symbol $\sum_{\{S_k \}}$ marks a summation over all possible spin configurations of the nodal Ising spins. After performing the latter two summations over spin states of two interstitial Ising spins $\mu_{k,1}$ and $\mu_{k,2}$ one gets the effective Boltzmann's factor, which can be subsequently replaced with a simpler equivalent expression through the generalized decoration-iteration transformation \cite{fis59,roj09,str10}
\begin{eqnarray}
{\cal Z}_k &=& \sum_{\mu_{k,1}} \sum_{\mu_{k,2}} \exp (-\beta {\cal H}_k)
= 2 \exp \left[ \frac{\beta h}{2} \left(S_{k} + S_{k+1} \right) \right] \nonumber \\
&& \times \left \{ \exp \left( - \frac{\beta J_2}{4} \right)
\cosh \left[\frac{\beta}{2} \left(J_1 + J_3 \right) \left(S_{k} + S_{k+1} \right) - \beta h \right] \right.
\nonumber \\
&& \left. \quad + \exp \left(\frac{\beta J_2}{4} \right)
\cosh \left[\frac{\beta}{2} \left(J_1 - J_3 \right) \left(S_{k} - S_{k+1} \right)\right] \right \}
\nonumber \\
&=& A \exp \left[\beta R S_k S_{k+1} + \frac{\beta D}{2} \left(S_k^2 + S_{k+1}^2 \right) + \beta Q S_k^2 S_{k+1}^2 \right.
\nonumber \\
&& \left. ~\qquad + \beta L \left(S_k S_{k+1}^2 + S_k^2 S_{k+1} \right) + \frac{\beta  h_0 }{2}\left(S_k + S_{k+1} \right) \right].
\label{dit}
\end{eqnarray}
As usual, the transformation parameters $A$, $R$, $D$, $Q$, $L$, and $h_0$ are given by a self-consistency condition of the decoration-iteration transformation (\ref{dit}), which requires that this mapping transformation must be valid independently of the spin states of two nodal Ising spins $S_k$ and $S_{k+1}$ involved therein \cite{fis59,roj09,str10}. By substituting all nine available spin states of two nodal Ising spins $S_k$ and $S_{k+1}$ into the transformation formula (\ref{dit})
one merely gets six different expressions for the effective Boltzmann's factor ${\cal Z}_{S_k, S_{k+1}}={\cal Z}_k (S_k, S_{k+1})$
\[
{\cal Z}_{0,0} = 2 \exp \left( -\frac{\beta J_2}{4} \right)
\cosh \left(\beta h \right) + 2 \exp \left(\frac{\beta J_2}{4} \right),
\]
\vspace{-3mm}
\begin{eqnarray*}
{\cal Z}_{ \pm 1,0} &{=}& 2 \exp \left( {\pm} \frac{\beta h}{2} \right)
\! \left \{ \exp \left( \frac{\beta J_2}{4} \right) \cosh \left[\frac{\beta}{2} \left(J_1 - J_3 \right) \right] \right.
\\
&& \qquad\qquad\quad
\left. {+} \exp \left( - \frac{\beta J_2}{4} \right)
\cosh \left[\frac{\beta}{2} \left(J_1 {+} J_3 \mp 2h \right) \right] \right \},
\end{eqnarray*}
\vspace{-3mm}
\begin{eqnarray*}
{\cal Z}_{1,1} {=} 2 \exp \left(\beta h \right) \! \left \{ \exp \left( \frac{\beta J_2}{4} \right)
{+} \exp\! \left(\! {-} \frac{\beta J_2}{4} \! \right) \cosh \left[\beta \left(J_1 {+} J_3 {-} h \right) \right] \right \},
\end{eqnarray*}
\vspace{-3mm}
\begin{eqnarray*}
{\cal Z}_{1,-1} {=} 2 \exp \left( \frac{\beta J_2}{4} \right) \cosh \left[\beta \left(J_1 {-} J_3 \right) \right]
{+} 2 \exp  \left( - \frac{\beta J_2}{4} \right) \cosh \left( \beta h \right),
\end{eqnarray*}
\vspace{-3mm}
\begin{eqnarray}
{\cal Z}_{-1,-1} &{=}& 2 \exp \left({-}\beta h \right) \left \{ \exp \left( \frac{\beta J_2}{4} \right) \right.
\nonumber\\
&& \qquad\qquad\quad
\left. {+} \exp \left( {-} \frac{\beta J_2}{4} \right) \cosh \left[\beta \left(J_1 {+} J_3 {+} h \right) \right] \right \},
\label{bf}
\end{eqnarray}
which unambiguously determine so far unspecified mapping parameters through the relations
\begin{eqnarray}
A = {\cal Z}_{0,0}, \quad
\beta R = \frac{1}{4} \ln \left(\frac{{\cal Z}_{1,1} {\cal Z}_{-1,-1}}{{\cal Z}^{2}_{1,-1}}\right), \nonumber
\end{eqnarray}
\vspace{-5mm}
\begin{eqnarray}
\beta D &=& \ln \left( \frac{{\cal Z}_{1,0} {\cal Z}_{-1,0}}{{\cal Z}^{2}_{0,0}}\right), \quad
\beta Q = \frac{1}{4} \ln \left( \frac{{\cal Z}_{1,1} {\cal Z}_{-1,-1}{\cal Z}^2_{1,-1}{\cal Z}_{0,0}^4}{{\cal Z}^4_{1,0}{\cal Z}^4_{-1,0}}\right),
\nonumber \\
\beta L &=& \frac{1}{4} \ln \left( \frac{{\cal Z}_{1,1} {\cal Z}^2_{-1,0}}{{\cal Z}_{-1,-1} {\cal Z}^2_{1,0}} \right),
\quad \beta h_0 = \ln \left( \frac{{\cal Z}_{1,0}}{{\cal Z}_{-1,0}}\right).
\label{mp}
\end{eqnarray}
If the decoration-iteration transformation (\ref{dit}) with the mapping parameters obeying the relations~(\ref{bf})-(\ref{mp}) is now substituted into Eq.~(\ref{pf})
one in turn obtains a simple mapping correspondence
\begin{eqnarray}
{\cal Z} (\beta, J_1, J_2, J_3, h) = A^N {\cal Z}_{\rm BEG} (\beta, R, D, Q, L, h_0),
\label{pf}
\end{eqnarray}
which relates the partition function ${\cal Z}$ of the mixed spin-(1,1/2) Ising diamond chain to the partition function ${\cal Z}_{\rm BEG}$ of the corresponding spin-1
Blume-Emery-Griffiths (BEG) chain \cite{blu71,kri74,kri75} given by the effective Hamiltonian
\begin{eqnarray}
{\cal H}_{\rm BEG} = &-& R \sum_{k=1}^{N} S_k S_{k+1} - D \sum_{k=1}^{N} S_k^2 - Q \sum_{k=1}^{N} S_k^2 S_{k+1}^2 \nonumber \\
                     &-& L \sum_{k=1}^{N} (S_k S_{k+1}^2 + S_k^2 S_{k+1}) - h_0 \sum_{k=1}^{N} S_k.
\label{beg}
\end{eqnarray}
It is quite evident from the effective Hamiltonian (\ref{beg}) that the mapping parameters $R$, $D$, $Q$, $L$, and $h_0$ represent the effective bilinear interaction, the single-ion anisotropy, the biquadratic interaction, the two-spin third-order interaction and the magnetic field of the corresponding spin-1 BEG chain.

The exact solution for the partition function of the generalized spin-1 BEG chain can easily be found by means of the transfer-matrix approach \cite{kri74,kri75,bax82}. Let us therefore merely recall the basic steps of this well-known procedure. The partition function of the spin-1 BEG chain can be first factorized into the following product
\begin{eqnarray}
{\cal Z}_{\rm BEG} = \sum_{S_1} \sum_{S_2} \cdots \sum_{S_N} \prod_{k=1}^N  {\rm T} (S_k, S_{k+1}),
\label{pfbeg}
\end{eqnarray}
where the expression ${\rm T} (S_k, S_{k+1})$ is defined as
\begin{eqnarray}
{\rm T} (S_k, S_{k+1}) &=& \exp \left[\beta R S_k S_{k+1}
+ \frac{\beta D}{2} \left(S_k^2 + S_{k+1}^2 \right) + \beta Q S_k^2 S_{k+1}^2 \right. \nonumber \\
&& \left. + \beta L \left(S_k S_{k+1}^2 + S_k^2 S_{k+1}\right) + \frac{\beta h_0}{2} \left(S_k {+} S_{k+1} \right)\right].
\label{tm}
\end{eqnarray}
The relevant expression given by Eq.~(\ref{tm}) can be considered as the usual transfer matrix
\begin{eqnarray}
{\rm T} (S_k, S_{k+1}) = \left( \begin{array}{ccc}
{\rm T} (1,1) & {\rm T} (1,0) & {\rm T} (1,-1) \\
{\rm T} (0,1) & {\rm T} (0,0) & {\rm T} (0,-1) \\
{\rm T} (-1,1) & {\rm T} (-1,0) & {\rm T} (-1,-1)
\end{array}
\right),
\nonumber
\end{eqnarray}
whereas the sequential summation over spin states of the spin-1 BEG chain will correspond to a multiplication of the relevant transfer matrices. As a result, the partition function can easily be calculated with the help of the respective eigenvalues $\lambda_i$ of the transfer matrix with regard to
\begin{eqnarray}
{\cal Z}_{\rm BEG} = \! \sum_{S_1 = \pm 1, 0} {\rm T}^N (S_1, S_1) = \mbox{Tr} \, {\rm T}^N  = \sum_{i=1}^3 \lambda_i^N.
\label{pfbege}
\end{eqnarray}
For completeness, let us quote the final expressions for the three transfer-matrix eigenvalues
\begin{eqnarray}
\lambda_i = r + 2 \, {\rm sgn} (q) \, \sqrt{p} \cos \left[\phi + (i-1) \frac{2\pi}{3} \right],
\label{evtm}
\end{eqnarray}
where
\begin{eqnarray}
{\rm sgn} (q) = \left \{\begin{array}{rl}
 -1, & ~q < 0
\\[3pt]
  1, & ~q \geq 0
\end{array}\right. , \nonumber
\end{eqnarray}
\vspace{-5mm}
\begin{eqnarray}
r &=& \frac{1}{3} \left[1 + 2 \exp(\beta R + \beta D + \beta Q) \cosh \left(\beta h_0 + 2 \beta L \right) \right], \nonumber \\
p &=& \frac{1}{4} (r-1)^2 + \frac{1}{3} \exp(2 \beta R + 2 \beta D + 2 \beta Q) \sinh^2 \left(\beta h_0 + 2 \beta L \right) \nonumber \\
&& + \frac{1}{3} \exp(-2 \beta R + 2 \beta D + 2 \beta Q) + \frac{2}{3} \exp(\beta D) \cosh \left(\beta h_0 \right), \nonumber \\
q &=& r^3 - \exp(\beta R {+} 2 \beta D {+} \beta Q) \cosh \left(2 \beta L \right)
      + \exp(-\beta R {+} 2 \beta D {+} \beta Q) \nonumber \\
&& + r \exp(\beta D) \cosh \left(\beta h_0 \right) + (1 {-} r) \exp(2 \beta D + 2 \beta Q) \sinh (2 \beta R) \nonumber \\
&& - r  \exp(\beta R + \beta D + \beta Q) \cosh (\beta h_0 + 2 \beta L),  \nonumber \\
\phi &=& \frac{1}{3} \arctan \left(\frac{\sqrt{p^3 - q^2}}{q} \right).
\label{pqr}
\end{eqnarray}
In the thermodynamic limit $N \to \infty$, the partition function as well as the associated free energy per site of the spin-1 BEG chain is simply given by the largest transfer-matrix eigenvalue $\lambda_{\rm max} = {\rm max} \{ \lambda_1, \lambda_2, \lambda_3 \}$ 
\begin{eqnarray}
f_{\rm BEG} = -\frac{1}{\beta} \lim_{N \to \infty} \frac{1}{N} \ln {\cal Z}_{\rm BEG} = -\frac{1}{\beta} \ln \lambda_{\rm max}.
\label{frebege}
\end{eqnarray}

Our rigorous calculation for the partition function of the mixed spin-(1,1/2) Ising diamond chain is thus formally completed, since it is now sufficient to substitute the exact result (\ref{pfbeg}) for the partition function of the corresponding spin-1 BEG chain into the relevant mapping relation (\ref{pf}) between both partition functions. As a result, the reduced free energy of the mixed spin-(1,1/2) Ising diamond chain per unit cell reads
\begin{eqnarray}
f = f_{\rm BEG} -\frac{1}{\beta} \ln A = -\frac{1}{\beta} (\ln A + \ln \lambda_{\rm max}),
\label{free}
\end{eqnarray}
while the total magnetization normalized with respect to its saturation value readily follows from the relation
\begin{eqnarray}
\frac{m}{m_s} = - \frac{1}{2} \frac{\partial f}{\partial h} = \frac{1}{2} \left[\frac{\partial \ln A}{\partial (\beta h)} 
              + \frac{\partial \ln \lambda_{\rm max}}{\partial (\beta h)}\right].
\label{mag}
\end{eqnarray}

\section{Results and discussions}
\label{result}

Now, let us proceed to a discussion of the most interesting results obtained for the mixed spin-(1,1/2) Ising diamond chain with the antiferromagnetic interactions 
($J_1, J_2, J_3>0$), whose magnetic behaviour may be fundamentally affected by the geometric frustration triggered by the competing diagonal interaction $J_2$ present in 
the diamond-like units. It is quite evident that the magnetic properties of the mixed spin-(1,1/2) Ising diamond chain given by the Hamiltonian (\ref{htot}) remain invariant under the transformation $(J_1, J_3) \to (J_3, J_1)$, which allows us to consider $J_1 \geq J_3$ without loss of generality. For simplicity, we will pass to the dimensionless parameters $J_2/J_1$, $J_3/J_1$, and $h/J_1$ by normalizing all the interaction parameters with respect to the exchange constant $J_1$, which will hereafter serve as the energy unit. While the former interaction parameter $J_2/J_1$ measures a degree of the geometric frustration inherent in the investigated mixed-spin diamond chain, the latter interaction parameter $J_3/J_1 \in [0,1]$ characterizes a degree of the asymmetry of both Ising couplings along sides of the diamond units.

\begin{figure*}
\begin{center}
\includegraphics[width=0.85\textwidth]{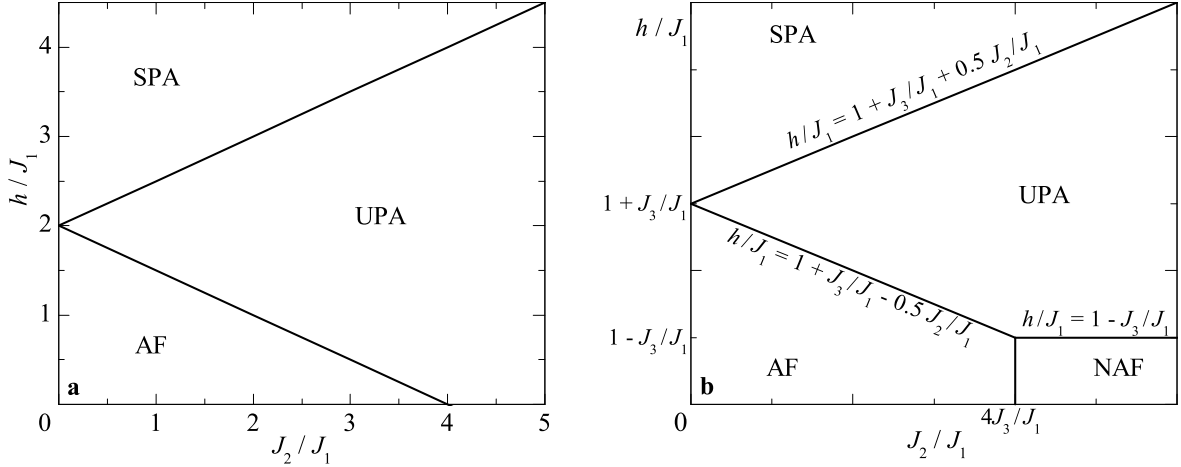}
\end{center}
\vspace{-0.8cm}
\caption{The ground-state phase diagram in the $J_2/J_1-h/J_1$ plane for: (a) the symmetric diamond chain with $J_3/J_1=1$; (b) the asymmetric diamond chain with $J_3/J_1<1$. 
The ground-state phase boundaries are explicitly quoted along the relevant lines in Fig.~\ref{fig2}b.}
\label{fig2}
\end{figure*}

First, let us establish the ground-state phase diagram for the symmetric as well as the asymmetric version of the mixed spin-(1,1/2) Ising diamond chain. Depending on the interplay between the interaction parameters $J_2/J_1$, $J_3/J_1$, and $h/J_1$, one finds in total four different ground states to be further referred to as the antiferromagnetic state (AF), the nodal antiferromagnetic state (NAF), the unsaturated paramagnetic state (UPA), and the saturated paramagnetic state (SPA). The relevant ground states are unambiguously given by the following spin configurations quoted along with the respective ground-state energies per primitive unit cell  
\begin{eqnarray}
|\mbox{AF} \rangle &=& \left \{\begin{array}{l}
\prod\limits_{k=1}^N \left|S_k = -1 \right\rangle \, \left|\mu_{k,1} =  \frac{1}{2} \right\rangle \, \left|\mu_{k,2} =  \frac{1}{2} \right\rangle
\\
\prod\limits_{k=1}^N \left|S_k =  1 \right\rangle \, \left|\mu_{k,1} = -\frac{1}{2} \right\rangle \, \left|\mu_{k,2} = -\frac{1}{2} \right\rangle
\end{array}\right.,
\nonumber \\
{\cal E}_{\rm{AF}}  &=& -J_1 + \frac{J_2}{4} - J_3,
\nonumber \\
|\mbox{NAF} \rangle &=& \left \{\begin{array}{l}
{\prod\limits_{k=1}^N \left|S_k = (-1)^{k} \right\rangle \,
\left|\mu_{k,1} = \frac{(-1)^{k+1}}{2} \right\rangle \, \left| \mu_{k,2} = \frac{(-1)^{k}}{2} \right\rangle}
\\
{\prod\limits_{k=1}^N \left|S_k = (-1)^{k+1} \right\rangle \,
\left|\mu_{k,1} = \frac{(-1)^{k}}{2} \right\rangle \, \left|\mu_{k,2} = \frac{(-1)^{k+1}}{2} \right\rangle}
\end{array}\right.,
\nonumber \\
{\cal E}_{\rm{NAF}} &=& -J_1 -\frac{J_2}{4} + J_3,
\nonumber \\
|\mbox{UPA} \rangle &=& \prod\limits_{k=1}^N \left|S_k = 1 \right\rangle \,
\left \{
\left|\mu_{k,1} =  \frac{1}{2} \right\rangle \, \left|\mu_{k,2} = -\frac{1}{2} \right\rangle
\atop
\left|\mu_{k,1} = -\frac{1}{2} \right\rangle \, \left|\mu_{k,2} =  \frac{1}{2} \right\rangle
\right.,
\nonumber \\
{\cal E}_{\rm{UPA}} &=& -\frac{J_2}{4} - h,
\nonumber \\
|\mbox{SPA} \rangle &=&  \prod\limits_{k=1}^N \left|S_k = 1 \right\rangle \, \left|\mu_{k,1} = \frac{1}{2} \right\rangle \, \left|\mu_{k,2} = \frac{1}{2} \right\rangle,
\nonumber \\
{\cal E}_{\rm{SPA}} &=&  J_1 + \frac{J_2}{4} + J_3 - 2 h.
\end{eqnarray}
Apparently, the two-fold degenerate AF ground state corresponds to the antiferromagnetic ordering, at which the nodal Ising spins are aligned in opposite to the interstitial Ising spins. Even though the nodal spins have a twice as large magnetic moment as the interstitial spins, the total magnetization completely cancels out on behalf of a twice as large number of the interstitial spins.\footnote{Note that both sublattice magnetizations completely cancel out just when assuming the equal $g$-factors for two different magnetic entities as originally imposed in the Hamiltonian (\ref{htot}). In the real magnetic substances, one should however expect at least some small difference between the relevant $g$-factors, which should result in a non-zero total magnetization and the ferrimagnetic rather than the antiferromagnetic spin arrangement.} 
Another two spin configurations inherent to the NAF ground state involve the antiferromagnetic alignment of the interstitial spins from the same diamond unit, 
as well as, the antiferromagnetic alignment of the nodal spins from the nearest-neighbour diamond units. Under this condition, one observes an interesting doubling of the magnetic unit cell when comparing it with the primitive diamond unit cell of the mixed-spin Ising diamond chain. However, the most intriguing spin alignment can be found in the macroscopically degenerate UPA ground state, where the nodal spins are fully polarized by the external magnetic field and the interstitial spins from the same diamond unit occupy one out of two equiprobable antiferromagnetic states $|\mu_{k,1} = 1/2 \rangle \, |\mu_{k,2} = -1/2 \rangle$ and $|\mu_{k,1} = -1/2 \rangle \, |\mu_{k,2} =1/2 \rangle$. The macroscopic degeneracy of the UPA ground state is accordingly proportional to a total number of the interstitial spin pairs, which is also reflected in the respective value of the residual entropy ${\cal S}_{\rm res} = N k_{\rm B} \ln2$. At sufficiently high magnetic fields, one finally detects the trivial SPA ground state with a full alignment of all nodal as well as interstitial spins into the external magnetic field.  

\begin{figure*}
\begin{center}
\includegraphics[width=0.85\textwidth]{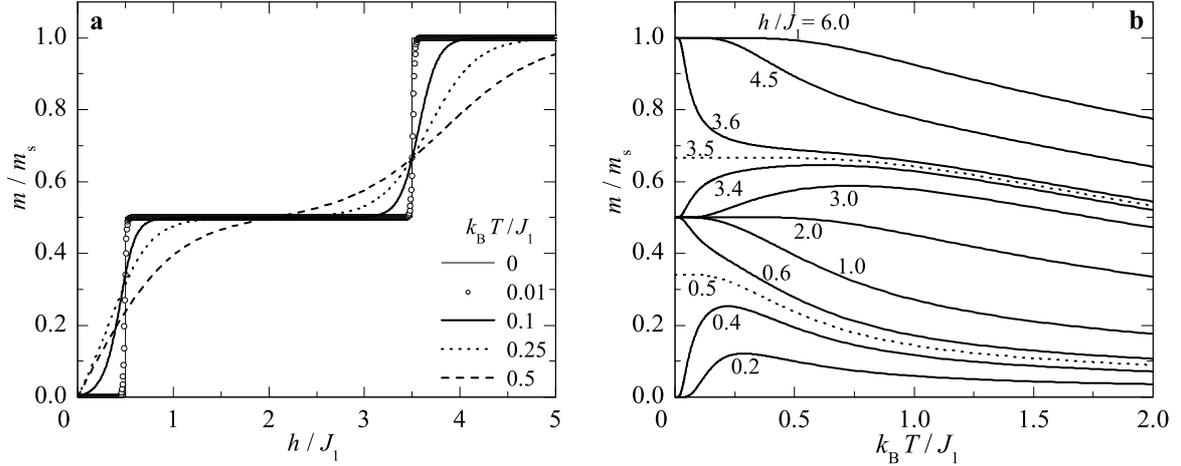}
\end{center}
\vspace{-0.8cm}
\caption{(a) The total magnetization normalized with respect to its saturation value as a function of the magnetic field for the asymmetric diamond chain with 
$J_2/J_1=4$, $J_3/J_1 = 0.5$ at a few different temperatures. Note that this choice of the parameters coincides with that one reported in Fig.4b of Ref.~\cite{xin12}; 
(b) Temperature variations of the total magnetization for the asymmetric diamond chain with $J_2/J_1=4$, $J_3/J_1 = 0.5$ at several values of the magnetic field.}
\label{fig3}
\end{figure*} 

The relevant ground-state phase diagram including all the available ground states is displayed in Fig.~\ref{fig2} for the particular case of the symmetric diamond chain (Fig.~\ref{fig2}a), as well as, the more general case of the asymmetric diamond chain (Fig.~\ref{fig2}b). Let us at first comment the ground-state phase diagram of the symmetric diamond spin chain. It is quite obvious from Fig.~\ref{fig2}a that the interaction term $J_2/J_1$, which is responsible for a geometric spin frustration, enhances a stability region of the UPA ground state, whereas there are two different scenarios of the magnetization process depending on whether $J_2/J_1 < 4$ or $J_2/J_1 \geq 4$. In the former case, the AF spin alignment is being the respective ground state at low enough fields, the UPA ground state develops in a range of moderate fields and finally, the SPA ground state is stabilized at sufficiently high fields. On the other hand, the AF spin arrangement does not exist in the ground state for the latter particular case with $J_2/J_1 \geq 4$. Under this condition, the macroscopically degenerate UPA spin arrangement always represent the respective ground state below the saturation field, which determines the field-induced transition towards the fully aligned SPA ground state. As far as the ground-state phase diagram of the more general asymmetric diamond chain (Fig.~\ref{fig2}b) is concerned, one observes the same general trends in the relevant ground-state phase diagram with exception of a presence of the additional possible ground state that corresponds to the NAF spin arrangement. It can be easily understood from Fig.~\ref{fig2}b that the existence of the NAF ground state is restricted to the parameter region, where $J_2/J_1 > 4 J_3/J_1$ and simultaneously $h/J_1 < 1 - J_3/J_1$. The former inequality implies that the NAF ground state replaces the AF one in a rather wide region of the parameter space when considering the highly asymmetric diamond chain with $J_3/J_1 << 1$, which is also the condition of a sufficiently high persistence of the NAF ground state with respect to the external magnetic field according to the latter inequality.

To provide an independent verification of the established ground-state phase diagram, let us examine in detail the magnetization process of the mixed spin-(1,1/2) Ising diamond chain by investigating the field dependence of the total magnetization at different temperatures. For illustration, Fig.~\ref{fig3}a depicts the total magnetization normalized with respect to its saturation value in dependence on the reduced magnetic field for the fixed values of the interaction parameters $J_2/J_1=4$ and $J_3/J_1 = 0.5$. It is noteworthy that the zero-temperature magnetization curve, which is in accordance with the ground-state phase diagram shown in Fig.~\ref{fig2}b, is plotted in Fig.~\ref{fig3}a by a thin solid line along with the finite-temperature magnetization curves calculated with the help of Eq.~(\ref{mag}). In agreement with the ground-state phase diagram (Fig.~\ref{fig2}b), the zero-temperature magnetization curve exhibits two abrupt magnetization jumps reflecting two consecutive field-induced transitions between NAF-UPA 
and UPA-SPA at the relevant transition fields $h_{\rm c1}/J_1 = 0.5$ and $h_{\rm c2}/J_1 = 3.5$, respectively. It can be clearly seen from Fig.~\ref{fig3}a that the low-temperature magnetization curves closely follow the displayed zero-temperature magnetization curve (see for instance the magnetization curve for $k_{\rm B} T/J_1 = 0.01$), which proves a correctness of the established ground-state phase diagram and disproves an existence of any further ground state that would be reflected in some additional magnetization plateau. Apparently,  the observed magnetization jumps and magnetization plateaus at zero and one-half of the saturation magnetization are just gradually smudged upon increasing temperature. To summarize, the low-temperature magnetization curves of the mixed spin-(1,1/2) Ising diamond chain exhibit at most two different magnetization plateaus before reaching the saturation magnetization, which coincide with a presence of the AF or NAF ground state with zero total magnetization and of the UPA ground state with the total magnetization equal to a half of the saturation magnetization.

Next, let us focus on typical temperature dependences of the total magnetization as depicted in Fig.~\ref{fig3}b for the same set of the interaction parameters $J_2/J_1=4$ and $J_3/J_1 = 0.5$ at several values of the external magnetic field. If the magnetic field is lower than the first critical field $h<h_{\rm c1}$, then, the total magnetization exhibits a round temperature-induced maximum as a function of the temperature when falling towards zero at sufficiently low and high temperatures. Contrary to this, the total magnetization starts from the one-half of the saturation value for the mediate magnetic fields $h \in (h_{\rm c1}, h_{\rm c2})$. The magnetization then either shows a monotonous temperature-induced decline for the magnetic fields slightly above the first transition field (i.e. $h \gtrapprox h_{\rm c1}$), or the non-monotonous temperature dependence 
with a round maximum for the magnetic field slightly below the saturation field (i.e. $h \lessapprox h_{\rm c2}$). Finally, the total magnetization always exhibits a more or less steep temperature-induced decrease when starting from the saturation value for the magnetic fields stronger than the saturation field $h>h_{\rm c2}$. For completeness, let us quote that the total magnetization normalized with respect to its saturation value starts from the non-trivial values $0.34151$ and $2/3$ for two particular magnetic fields equal to the transition fields $h_{\rm c1}$ and $h_{\rm c2}$, respectively.

Last but not least, let us make a few comments on the magnetization curves of the mixed spin-(1,1/2) Ising diamond chain obtained by Zihua Xin \textit{et al}. \cite{xin12} by employing the Monte Carlo simulations. Note that Xin and co-workers have reported in Ref. \cite{xin12} two additional intermediate magnetization plateaus at 0.283 and 0.426 
of the saturation magnetization, which were ascribed to a presence of some metastable states in the low-temperature magnetization curves. It should be pointed out, however, that the standard Monte Carlo simulation based on the Metropolis algorithm has been employed in Ref. \cite{xin12}, which should only give the stable states in thermal equilibrium rather than metastable states. From this perspective, one has to refute both striking magnetization plateaus at 0.283 and 0.426 of the saturation magnetization reported in Ref. \cite{xin12}, since they evidently contradict the exact analytical results presented in this work. It is worthwhile to remark that Monte Carlo simulations were performed in Ref. \cite{xin12} at the unusually low temperature $k_{\rm B} T/J_1 = 0.0001$, which might indicate extremely long relaxation times needed for establishing the thermal equilibrium and hence, one may consider an insufficient number of Monte Carlo steps used for equilibration at a given very low temperature as the main reason for the above mentioned discrepancy.

\section{Conclusion}
\label{conclusion}

In the present article, the ground state and magnetization process of the mixed spin-(1,1/2) Ising diamond chain have been rigorously studied by combining the generalized decoration-iteration transformation with the transfer-matrix method. In particular, our attention was focused on possible magnetization scenarios leading to the intermediate magnetization plateaus and the overall nature of available ground states. It has been demonstrated that the ground-state phase diagram of the symmetric diamond chain totally consist of three different ground states, while the ground-state phase diagram of the more general asymmetric diamond chain includes in total four different ground states. It has been evidenced that two different magnetization scenarios may in principle occur for the symmetric as well as asymmetric mixed-spin diamond chain depending on a mutual interplay between the interaction parameters. It has been actually evidenced that the low-temperature magnetization curves of the symmetric and asymmetric diamond chains display at most two different magnetization plateaus, which manifest a presence of the AF or NAF ground states with zero total magnetization and/or UPA ground state with the total magnetization equal to a half of the saturation magnetization. Owing to this fact, our exact analytical calculations refute recent Monte Carlo simulations by Zihua Xin \textit{et al}. \cite{xin12}, which have predicted two additional striking magnetization plateaus at 0.283 and 0.426 of the saturation magnetization.

From the methodological point of view, we have adapted in the present work the generalized decoration-iteration transformation in order to establish a rigorous mapping correspondence of the investigated mixed spin-(1,1/2) Ising diamond chain with the effective spin-1 BEG chain. To the best of our knowledge, this form of the generalized mapping transformation has been adapted so far just for a calculation of the zero-field properties of the mixed spin-(1,3/2) Ising linear chain \cite{fir97,fir03}. The present work thus brings a rather simple recipe to greatly extend this rigorous mapping approach, which may be further substantially generalized in order to investigate magnetic properties of several exactly soluble mixed spin-(1,$S$) Ising and Ising-Heisenberg diamond chains in a non-zero magnetic field. As a matter of fact, the present approach can be rather straightforwardly extended in order to account for the quite general spin numbers of the interstitial spins, the more general Heisenberg interaction between the interstitial spins, the single-ion anisotropy, the next-nearest-neighbour and/or biquadratic interaction between the nodal spins and so on.

\end{document}